\documentclass{IEEEtran}
\usepackage{cite}
\usepackage{amsmath,amssymb,amsfonts}
\usepackage{graphicx}
\usepackage{stfloats}
\usepackage{textcomp,nicefrac}
\usepackage{lineno}
\usepackage{wrapfig}

\def\BibTeX{{\rm B\kern-.05em{\sc i\kern-.025em b}\kern-.08em
		T\kern-.1667em\lower.7ex\hbox{E}\kern-.125emX}}
\begin{document}
	\title{Redesign of the ATLAS Tile Calorimeter link Daughterboard for the HL-LHC.}
	
	\author{Eduardo Valdes Santurio, Samuel Silverstein, Christian Bohm \IEEEmembership{Senior, IEEE},\\ Katherine Dunne, Suhyun Lee, Holger Motzkau.	
		\thanks{Eduardo Valdes Santurio, Fysikum, Stockholm University, email: eduardo.valdes@fysik.su.se, eduardo.valdes@cern.ch. Samuel Silverstein, Fysikum, Stockholm University, email: silver@fysik.su.se. Christian Bohm , Fysikum, Stockholm University, email: bohm@fysik.su.se. Katherine Dunne, Fysikum, Stockholm University, email: katherine.dunne@fysik.su.se. Suhyun Lee, Fysikum, Stockholm University, email: suhyun.lee@fysik.su.se. Holger Motzkau, Fysikum, Stockholm University, email: holger.motzkau@fysik.su.se.
	\\
	This work was supported by Stockholm University and CERN.\\
	\textbf{Copyright 2020, CERN, for the benefit of the ATLAS Collaboration. \\CC-BY-4.0 license.}}}

	
	\maketitle
	
	\begin{abstract}
	The Phase-2 ATLAS upgrade for the High Luminosity Large Hadron Collider (HL-LHC) has motivated progressive redesigns of the ATLAS Tile Calorimeter (TileCal) read-out link and control board (Daughterboard). The Daughterboard (DB) communicates with the off-detector electronics via two 4.6 Gbps downlinks and two pairs of 9.6 Gbps uplinks. Configuration commands and LHC timing is received through the downlinks by two CERN radiation hard GBTx ASICs and propagated through Ultrascale+ FPGAs to the front-end. Simultaneously, the FPGAs send continuous high-speed readout of digitized PMT samples, slow control and monitoring data through the uplink. The design minimizes single points of failure and reduces sensitivity to SEUs and radiation damage by employing a double-redundant scheme, using Triple Mode Redundancy (TMR) and Xilinx Soft Error Mitigation (SEM) in the FPGAs, adopting Cyclic Redundancy Check (CRC) error verification in the uplinks and Forward Error Correction (FEC) in the downlinks. We present a DB redesign that brings an enhanced timing scheme, and improved radiation tolerance by mitigating Single Event Latch-up (SEL) induced errors and implementing a more robust power-up and current monitoring scheme.
	\end{abstract}
	
	
	\section{Introduction}
	The instantaneous luminosity at the HL-LHC will be increased by a factor of five compared to the LHC. Consequently, the read-out systems of the ATLAS detector\cite{bib_atlas} (Figure \ref{fig:atlas_tilecal}a) will be exposed to higher radiation levels and increased rates of pileup. The current TileCal read-out electronics will not be able to handle the new requirements imposed by the HL-LHC. R\&D work is ongoing to upgrade the TileCal electronics with a new design that will provide continuous digital read-out of all the calorimeter cells with lower electronic noise and better timing stability. The upgraded electronics will achieve better energy resolution and less sensitivity to out-of-time pileup \cite{bib_tilecal_phase_ii_tdr:2017}. The R\&D work requires Total Ionizing Dose (TID), Non Ionizing Energy Loss (NIEL), and Single Event Effects (SEE) tests to be performed on the upgraded on-detector electronics, to qualify the design as reliable for the HL-LHC radiation environment.

	\section{ATLAS Tile Calorimeter}
\begin{figure}[!ht]
	\centering
	\includegraphics[width=1.0\linewidth]{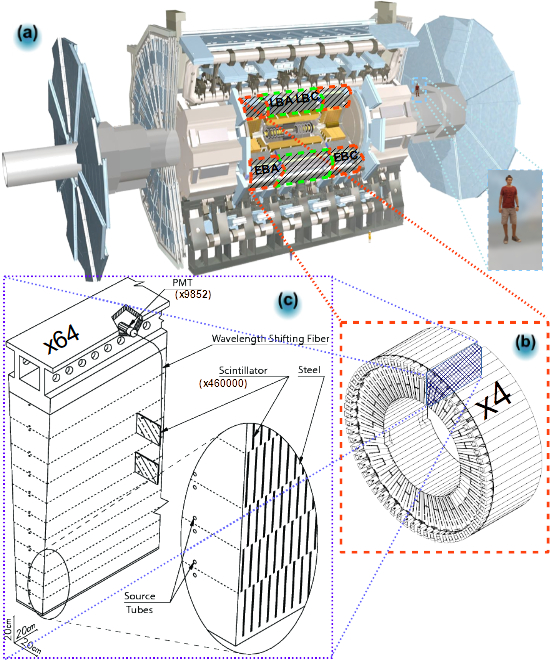}
	\caption{(a) The ATLAS detector. (b) A TileCal barrel. (c) Depiction of a TileCal wedge-shaped module.}
	\label{fig:atlas_tilecal}
\end{figure}

	TileCal is a sampling calorimeter with plastic scintillator tiles and steel plates as active and absorber materials, respectively. TileCal is longitudinally divided into three cylindrical barrels (Figure \ref{fig:atlas_tilecal}b) each comprising $64$ wedge-shaped modules (Figure \ref{fig:atlas_tilecal}c). Scintillator light is collected on each side of a pseudo-projective cell by wavelength shifting fibers and read out by a pair of PMTs.
	
\section{The TileCal HL-LHC read-out system}
\begin{figure*}[!ht]
	\centering
	\includegraphics[width=1.0\linewidth]{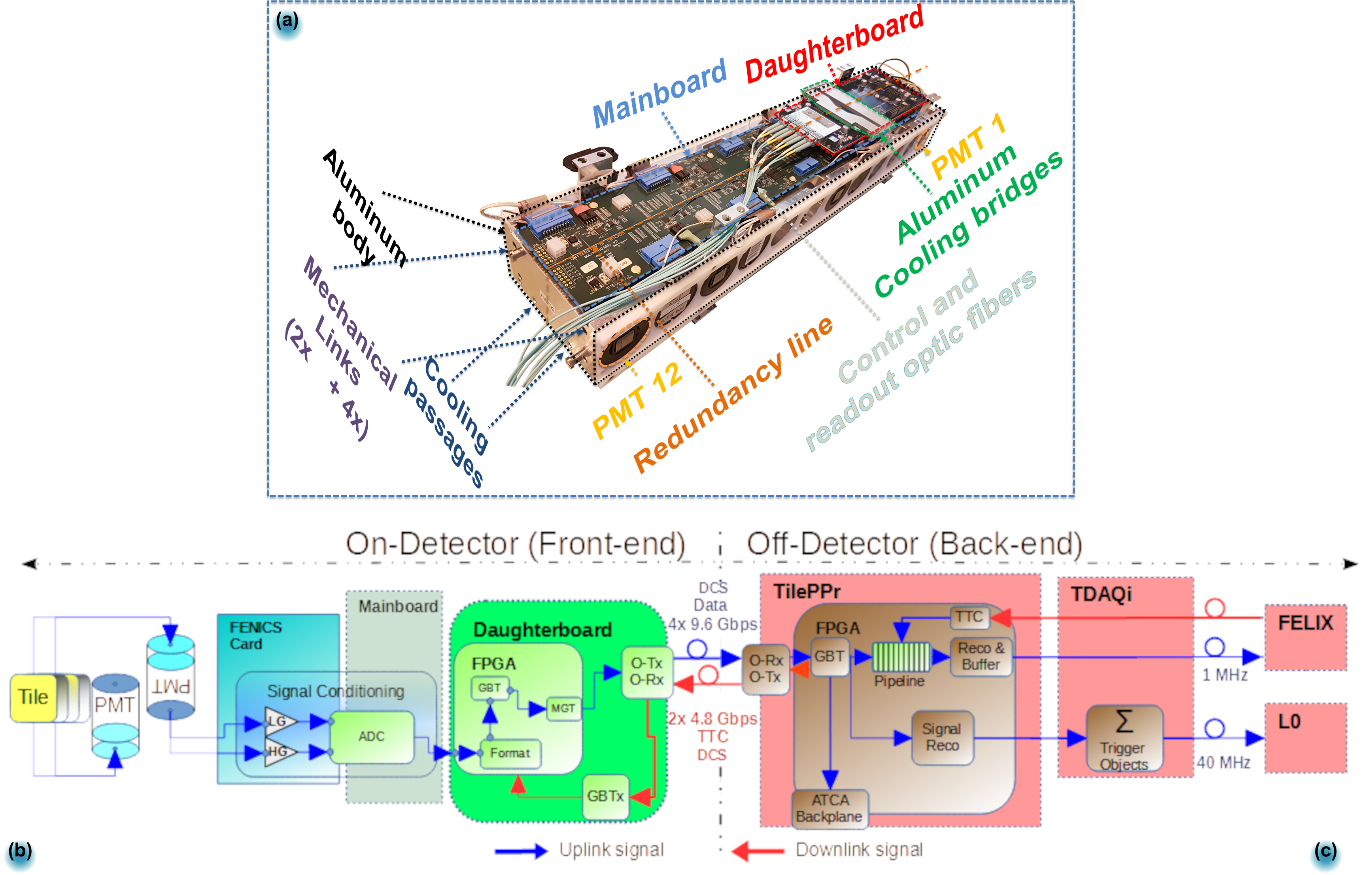}
	\caption{TileCal HL-LHC Upgrade read-out system: (a) TileCal Phase-II Upgrade Minidrawer, (b) On-Detector electronics block diagram and (c) Off-Detector electronics block diagram.}
	\label{fig:tilecal_md_phase_ii}
\end{figure*}

The TileCal Phase-II read-out on-detector electronics are mechanically subdivided into 896 independent modules, so-called Minidrawers (MD, Figure \ref{fig:tilecal_md_phase_ii}a) each servicing up to 12 PMT channels. Each PMT analogue signal is conditioned, shaped, and fed into amplifiers with low- and high-gain by a Front
End board for the New Infrastructure with Calibration and signal Shaping (FENICS) card, before they are digitized on a Mainboard (MB, Figure \ref{fig:tilecal_md_phase_ii}b). 

A MB link daughter board, called the Daughterboard (DB), receives and propagates LHC synchronized timing, control signals, and configuration commands to the front-end, while transmitting continuous read-out of the MB channels to the off-detector systems (Figure \ref{fig:tilecal_md_phase_ii}c). Tile Preprocessors (TilePPr) continuously receive and store PMT data in pipelines until reception of a trigger decision event, while providing processed and reconstructed data to the trigger system.

\section{The Daughterboard revision 5}
\begin{figure}[!ht]
	\centering
	\includegraphics[width=1.00\linewidth]{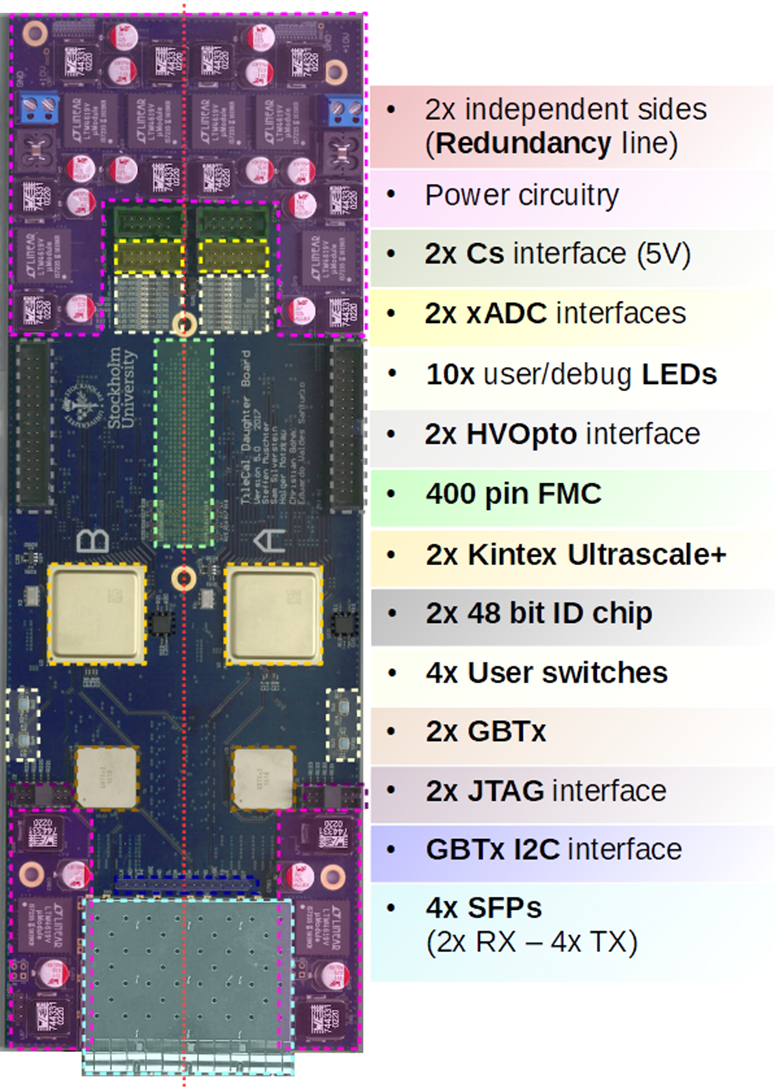}
	\caption{The Daughterboard revision 5.}
	\label{fig:db5}
\end{figure}

The DB revision 5 (DB5, Figure \ref{fig:db5}) has a double redundant design, allowing nominal running with two working links. Migration from the GTX older gigabit transceiver block present in the Kintex-7 Field Programmable Gate Arrays (FPGAs) in previous designs to the newer GTY gigabit transceivers present in the Kintex Ultrascale+ FPGAs, allows 9.6 Gbps communication with full compatibility with the LHC timing\cite{bib_db5}. Furthermore, improved radiation tolerance is expected, taking into consideration the lower cross-section for SEUs (Single Event Upsets) present in the Kintex Ultrascale+ with 16 nm FinFET technology compared to the Kintex 7 with 28\,nm TMSC process\cite{bib_seu_planar_finfet}. 

The design includes backwards compatibility with the interfaces of earlier versions, such as the connector to the on-detector high voltage interface (HV-OPTO), the MB FPGA Mezzanine Card (FMC) docking connector and the Caesium system connector (Cs Interface). The DB5 also includes a custom connector routed to each FPGAs' Xilinx Analogue to Digital Converter (xADC) for extra sensor signal monitoring, and new digital ID serial chips for unique digital identification of the MD. DB5 migrated from two QSFP optoelectric modules on the earlier design to four SFP+ optoelectric modules. 
Two SFP+ modules serve two redundant 4.8 Gbps input links to a pair of CERN custom radiation tolerant ASIC (GBTx)\cite{bib_gbtx_manual}. Each GBTx ASIC recovers two 160 MHz good quality LHC synchronized clocks in order to drive the transceivers of both FPGAs. Additionally, two 40 MHz Trigger, Timing and Control (TTC) synchronous clocks are recovered to drive the FPGAs relevant logic, and four 40 MHz TTC phase-configurable clocks to drive the digitizing blocks of each MB quadrant. Furthermore, each GBTx propagates configuration and control commands to both FPGAs through the dedicated GBTx ports (E-Ports), while providing remote control of the FPGA resets and JTAG chains for remote reconfiguration of the FPGAs and their attached PROMs. The two gains of digitized PMT data and Detector Control System (DCS) information are transmitted to the back-end by means of a pair of redundant 9.6 Gbps read-out uplinks from each FPGA, powered by TMR capable firmware providing continuous GBT-CRC protected words.

DB5 firmware has been under development and extensive testing has taken place in different testbenches connected to various prototype versions of the HL-LHC system. Important functionalities have been successfully implemented in the firmware. Stable 4.8 Gbps downlinks and 9.6 Gbps uplinks with the TilePPr were achieved. However, full performance of the uplinks was limited by jitter present in the GBTx de-skew clocks used to drive the GTY transceivers. DB5 was interfaced with different revisions of the MB, and the ADC read-out was successfully implemented with restricted performance because of the non-optimal routing of ADC signals and clocks to the FPGA I/O banks at the board level. Long runs reading-out the signals of the FENICS cards were performed to verify the functionalities of the whole read-out chain and the full system integration. The experience accumulated from the DB5 prototypes are a baseline guide for the new DB6 that is currently in active development.

\section{Radiation tests on the Daughterboard 5}
The DB5 was tested for TID, NIEL and SEE to verify the readiness of the design for the HL-LHC era. The design proved to withstand up to $20$\,kRad of TID, being enough to qualify it for the HL-LHC lifetime using the simulated radiation doses reported in the ATLAS Tile Calorimeter Phase-II Upgrade Technical Design Report\cite{bib_tilecal_phase_ii_tdr:2017}. However, new dose estimates were obtained by new simulations with updated geometry. The new values dictate that to qualify the board, a TID test should be done up to either $72$ kRad or to $14.4$ kRad if the test includes annealing. 

The NIEL test was performed with a $52$\,MeV proton beam, to deliver a total NIEL dose corresponding to $9.00\times 10^{12}$ neutrons$\times$cm$^{-2}$ 1\,MeV equivalent neutron fluence. Two different SFP+ devices (AVAGO AFBR-709SMZ and the CORETEK CT-000NPP-SB1L-D) were tested. After irradiation, the micro-controller in the AVAGO SFP+ completely lost functionality, with no signal or power consumption detected when connected. However the irradiated CORETEK SFP+ transceiver continued to function with no detected signal integrity degradation. The DB GBTx ASICs and the FPGAs were found to have been severely damaged, with the various supply voltages VCC shorted to ground. The target fluence was achieved at the expense of exposing the board to $768.2$\,kRad of equivalent TID. Therefore, the test was considered inconclusive and needs to be re-done in a facility capable of achieving the target fluence with neutrons.

\begin{figure}[!ht]
	\centering
	\includegraphics[width=0.90\linewidth]{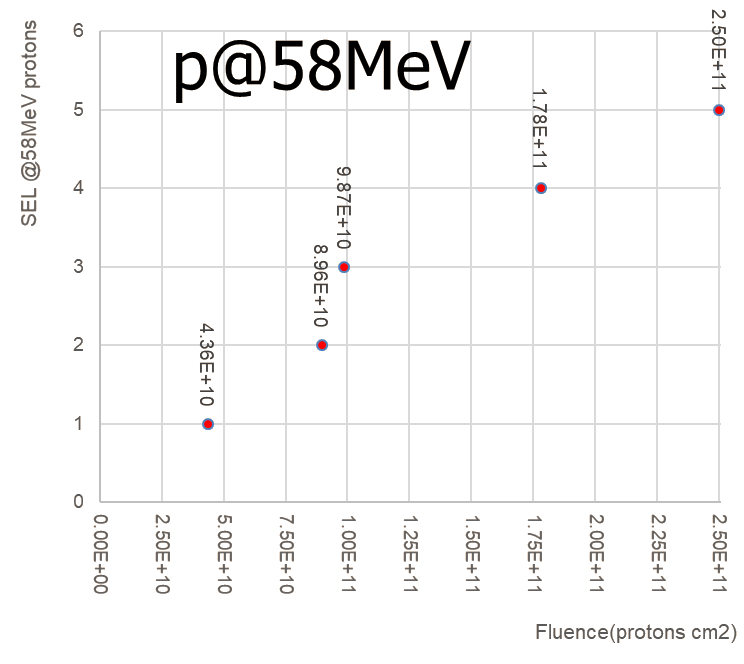}
	\caption{Single Event Latchups test for the Daughterboard 5. Test performed with protons at 58 MeV.}
	\label{fig:sel58mev}
\end{figure}

The SEE tests showed that the Kintex Ultrascale+ $16$ nm FinFET technology is susceptible to Single Event Latchups (SELs) with SEL-fluence rate increased from $2\times 10^{-11}$ SEL-fluence rate at $58$\,MeV (Figure \ref{fig:sel58mev}) to $2.36\times10^{-10}$ rate at $226$\,MeV protons. 

\begin{figure*}[t]
	\centering
	\includegraphics[width=0.8\linewidth, height=5cm]{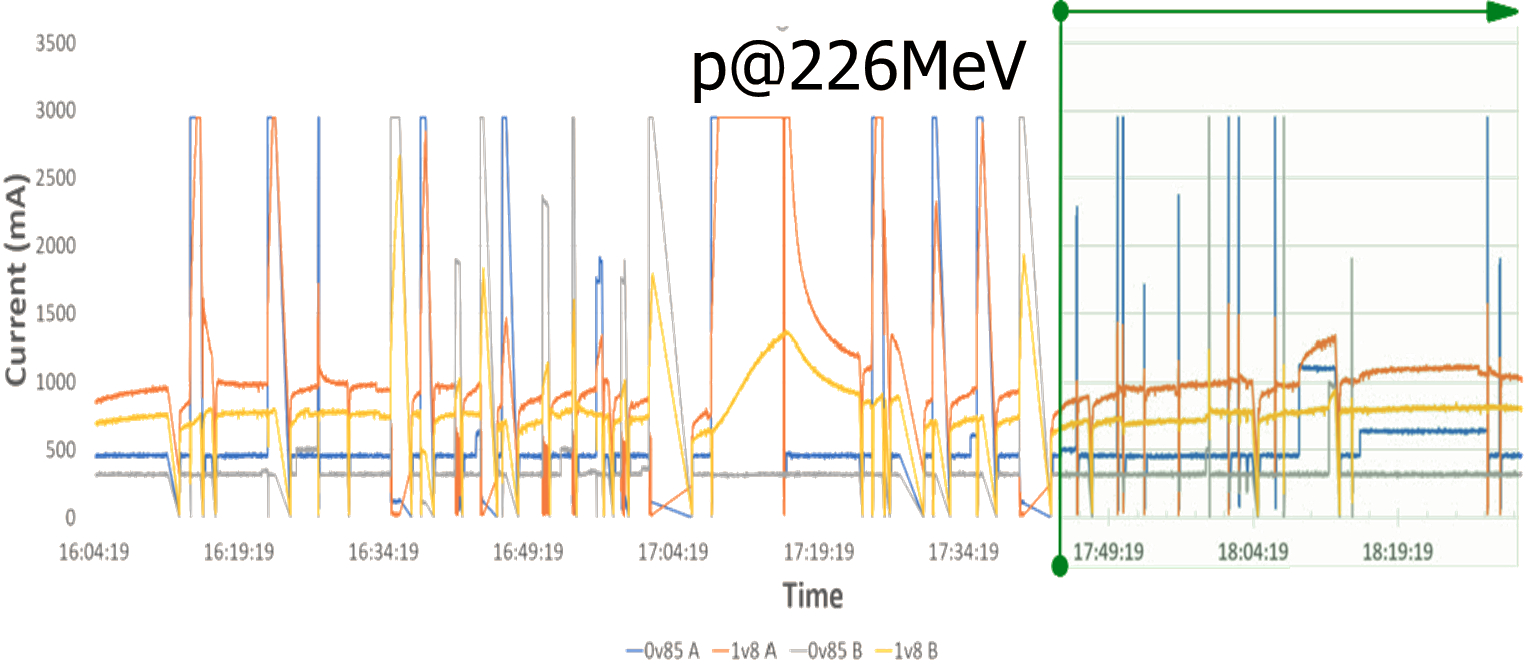}
	\caption{Current monitoring data corresponding to both irradiated FPGAs on the Single Event Latchups test performed with protons at 226 MeV for the Daughterboard 5 while SELs occurred. The highlighted part to the right of the chart corresponds to the time-frame where the power cycle was triggered within a second of the SEL event occurrence.}
	\label{fig:sel226mev}
\end{figure*}

Figure \ref{fig:sel226mev} shows the current monitored during the SEL test performed with a 226 MeV proton beam. Although the device under test was subjected to over-currents for relatively long periods of around 60\,s, the over-currents due to the latch-ups did not seem to cause any noticeable damage to any of the functionalities of the FPGAs tested. However, hardware susceptible to SELs is highly undesirable. Hence, the solutions to mitigate SELs include adding an over-current protection to the design and migrating to an FPGA that is not susceptible to SELs. The SEU tests resulted in $4934$ SEUs over a fluence of $2.05 \times 10^{11}$ protons cm$^{-2}$, of which only $11$ SEUs could not be corrected by the Xilinx Soft Error Mitigation (SEM) core. 

The SEU tests resulted in $4934$ SEUs over a fluence of $2.05 \times 10^{11}$ protons cm$^{-2}$, of which only $11$ SEUs could not be corrected by the Xilinx SEM. The uncorrectable SEU rates predict that 1.4 $\pm0.4$ uncorrectable errors are foreseen per Daughterboard per year. Between the Xilinx SEM and the TMR it is expected that correctable SEUs will not affect nominal runs.

\section{Redesign of the Daughterboard}

\begin{figure}[h]
	\centering
	\includegraphics[width=0.8\linewidth]{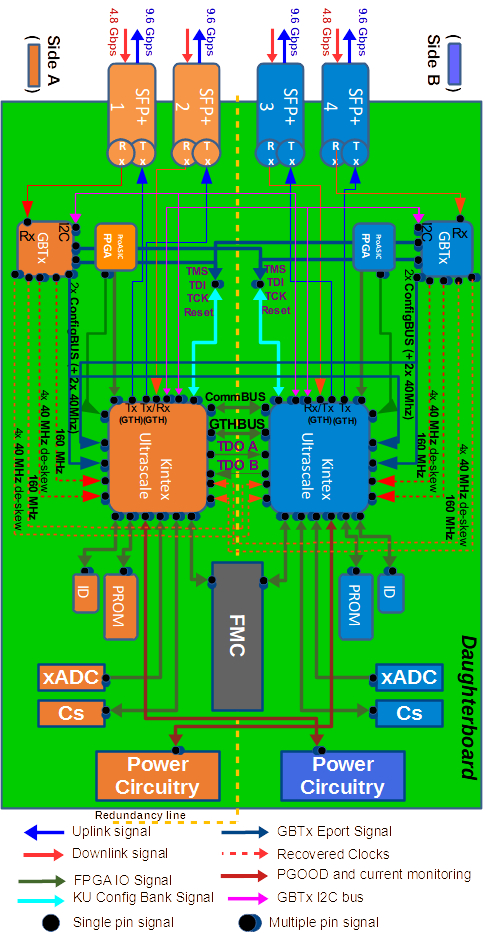}
	\caption{The Daughterboard revision 6 block diagram.}
	\label{fig:db6}
\end{figure}

The redesign of the DB aims to solve the radiation issues and incorporate a more robust timing design and extra features that will allow minimizing single points of failure one step further. However, backwards compatibility with previous interfaces will be maintained (Figure \ref{fig:db6}). 

DB6 will have the capability of using the four downlinks. The new design uses a Kintex Ultrascale XCKU035 FPGA powered by multiple GTH transceivers for high speed bandwidth communication. On each side, one SFP+ RX will be connected to the GBTx while the other will be connected to an FPGA GTH RX to test high-speed signal-quality. Additionally, two buses will allow communication between both FPGAs: a slow bus (CommBUS) and a high speed 9.6 Gbps bus powered by interconnected GTH transceivers (GTHBUS). The CommBUS and the GTHBUS interfaces will allow signal monitoring, internal trigger commands, and data interchange between both FPGAs.

\begin{figure*}[!ht]
	\centering
	\includegraphics[width=1.02\linewidth]{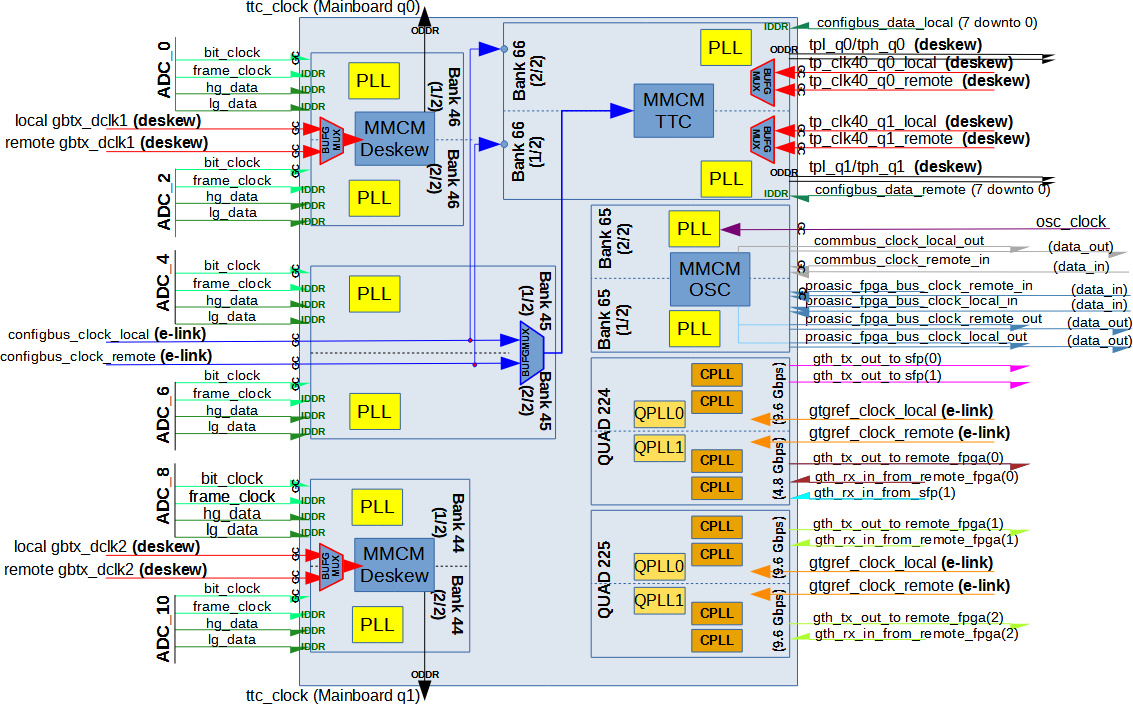}	\caption{Preliminary placement of clocks, ADC read-out pins, GTH signals, configbus interface and other clock dependent modules for XCKU035 FPGA to be used in DB6.} \label{fig_daughterboard_r6_sketch}
\end{figure*}

Migrating to a SEL-resistant FPGA is needed. The Ultrascale family is powered by a $20$ nm TSMC planar technology, where SELs have not been observed\cite{bib_sel_planar_finfet}. It needs to be taken into consideration that migrating to this FPGA family will increase the SEU rate by a factor of approximately $16$\cite{bib_seu_planar_finfet}. However, this level of increase on the SEU rates is well within the capabilities of SEM and TMR to mitigate.
Xilinx FPGAs require a specific power-on sequence and the implementation in DB5 was done by means of an RC circuit. This solution can be sensitive to a different number of parameters such as ageing, power ramping, or failure of one of the multiple LTM4619 DC-DC converters used in the board. The new design mitigates this issue by implementing a circuit that chains the DC-DC converters, taking advantage of the availability of the Power-Good signals to switch on each of the voltages in the recommended order. This solution will be integrated with a current limiting circuitry that will power off the DC-DC converter chain if one of the voltages gets a high current drain caused by a latch up, powering off the whole side affected in case of any voltage failure or over-current in any of the chain links. This solution will increase the robustness of the board by avoiding potential damages to the FPGAs and other components due to the presence of isolated unexpected over-currents caused by SELs or failures in the board power circuitry.

The DB6 target FPGA chosen is the XCKU035 from the Ultrascale family, that offers GTH transceivers fully compatible with the LHC timing and with very similar capabilities to the currently used XCKUP3 from the Ultrascale+ family. Additional radiation induced problems potentially caused by the use of large electrolytic capacitors in the DC-DC regulators will be mitigated by the use of conductive polymer capacitors, which have been observed to be radiation tolerant with no impact on performance after $200$ kRad at $500$ rad per hour\cite{bib_hard_rad_capacitors}.

The interfaces of the GBTxs with the JTAG chains of both FPGAs will be redesigned to mitigate misbehavior produced when the downlink is disconnected or becomes unstable. This event affects the output state signals RX\_RDY and DATA\_VALID, used to activate LVDS buffers dedicated to control the JTAG interfaces of both FPGAs via the E-Ports. It was found that digital noise propagates to the FPGA JTAG interface signals via the E-Ports which can cause the FPGAs to malfunction. The planned solution will be to buffer all of the signals of the JTAG interfaces through flash based Microsemi ProASIC3 FPGAs to enable or disable the selected interface by a sequence of commands from the GBTx.

The Kintex FPGAs are in charge of reading out the dual-gain PMT digitized data and formatting it into GBT words. The DB5 firmware has a complex clock and timing scheme with multiple clock inputs, from various sources such as the two GBTxs, an oscillator and the six ADCs sitting on a MB side. The DB5 design was found to have a sub-optimal routing that caused avoidable timing issues during the implementation of the firmware. The design proposal for DB6 includes taking advantage of the Ultrascale dedicated XYPHY BITSLICE byte groups architecture\cite{bib_ultrascale_clocking}. The bit-clock, the frame clock, the high-gain serial data, and the low-gain serial data signals coming from each ADC channel will be routed to the same byte XYPHY BITSLICE group. The byte group for each specific channel should be either 1 or 2, since they contain the dedicated Global Clock Capable IO pins (GCIO) that allow direct access to the Phase Locked Loops (PLLs) and the Mixed-Mode Clock Manager (MMCM) of the bank. Using a byte group per channel assures that there is no congestion with more than 6 clocks driving IO loads on any half bank. Figure\,\ref{fig_daughterboard_r6_sketch} shows a preliminary sketch for routing of the ADC clocks and relevant time-dependent signals to the different banks for the DB6 target FPGA. A clean timing-optimized firmware will be easier to implement with successful timing-closure.




\section{Conclusions}
The Daughterboard is the read-out link and control board that interfaces a TileCal Minidrawer and the off-detector electronics. The current revision of the Daughterboard (DB5) cannot entirely fulfill the radiation requirements imposed by the HL-LHC because of the Kintex Ultrascale+ 16 nm FinFET process sensitivity to SELs, and the design being incapable to mitigate their effects. The redesign will not only migrate to a more SEL-resistant FPGA, but also will add extra protection to SEL and better radiation tolerance by using conductive polymer capacitors. It is expected that the DB6 will succeed to meet the HL-LHC radiation requirements; however, it must still be tested and qualified for TID, NIEL and SEE as mandated by ATLAS policies. 

Additionally, the new design will further improve and optimize the already complex DB clocking scheme and bring an extra layer of hardware control and monitoring that will minimize single points of failure and allow mitigation of any unexpected issues during nominal runs. It is planned that approximately $1000$ of the redesigned Daughterboards will be produced for Phase-II as the contribution of Stockholm University to the ATLAS Upgrade for the HL-LHC era.

	\section*{References}
	
	\def\refname{\vadjust{\vspace*{-1em}}} 


\begin{thebibliography}{9} 
	\bibitem{bib_atlas}
	ATLAS Collaboration, 2008 JINST 3 S08003.
	
	\bibitem{bib_tilecal_phase_ii_tdr:2017}
	ATLAS collaboration. Technical Design Report for the Phase-II Upgrade of the ATLAS Tile Calorimeter. CERN-LHCC-2017-019, ATLAS-TDR-028, 2018.
	\bibitem{bib_db5}
	ATLAS Tile Calorimeter Link Daughterboard, PoS Volume 343 - Topical Workshop on Electronics for Particle Physics (TWEPP2018), DOI: 10.22323/1.343.0024.
	
	\bibitem{bib_seu_planar_finfet}
	Single Event Upset Characterization of the Zynq UltraScale+ MPSoC Using Proton Irradiation, 2017 IEEE Radiation Effects Data Workshop (REDW), DOI: 10.1109/NSREC.2017.8115448.
	
	\bibitem{bib_sel_planar_finfet}
	Single-Event Latch-Up: Increased Sensitivity From Planar to FinFET, 2017 IEEE Radiation, IEEE Transactions on Nuclear Science 2018, DOI: 10.1109/TNS.2017.2779831.
	
	\bibitem{bib_hard_rad_capacitors}
	Capacitors for Spacecraft: Withstanding a Harsh Radiation Environment, https://www.ttiinc.com/content/ttiinc/en/resources/marketeye/categories\\/passives/me-slovick-20160809.html.
	
	
	\bibitem{bib_ultrascale_clocking}
	UltraScale Architecture Clocking Resources User Guide.
	https://www.xilinx.com/support/documentation/user\_guides/ug572-ultrascale-clocking.pdf.	
	

	\bibitem{bib_gbtx_manual}
	GBTx Manual, CERN, 2018.
	https://espace.cern.ch/GBT-Project/GBTX/\\Manuals/gbtxManual.pdf.	
	
	
	
	
	
\end{thebibliography}
\end{document}